# Effect of Interlayer Shear to Graphene Resonators

Yilun Liu, Zhiping Xu, and Quanshui Zheng[*]

Department of Engineering Mechanics, CNMM, Tsinghua University, Beijing 100084, China

[*]Correspondence should be sent to zhengqs@tsinghua.edu.cn

**Abstract:**

Graphene nanostrips with single or a few layers can be made into bending resonators with extremely high sensitivity to environment changes. In this work we study the effect of interlayer shear on resonant frequencies $f$ of graphene nanostrips, via both molecular dynamics (MD) simulation and elastic model analysis incorporating interlayer shear. Contrary to the classical thin beam theory prediction $f \propto nl^{-2}$ ($l$ is beam length and $n$ layer number), MD simulation results reveal very different dependences, $f \propto l^{-1.36}$ and $f - f_{mono} \propto (n-1)/n$ ($f_{mono}$ is frequency of the monolayer beam). Interlayer shear modulus of multilayer graphene strips is much smaller than their intralayer Young's modulus, and the weak interlayer interaction can not maintain the registry between the carbon atoms in adjacent layers. Large shear deformation occurs during vibration of multilayer graphene nano-strips. Therefore we propose a multi-beam shear model (MBSM) with the interlayer shear energy of multilayer graphene nano-strips taken into account. It makes predictions consistent excellently with direct MD simulations without any fitting parameter required. The results are of importance for various applications of multi-layer graphene nano-strips, such as in nano-electromechanical devices including resonators, sensors and actuators, where interlayer shear has apparent impacts to deformation, vibration, and energy dissipation processes.

**Keywords:** multilayer graphene nano-strips, resonant frequency, interlayer shear model, nano-devices

# 1. Introduction

Two-dimensional graphene sheets, bonding together through in-plane $sp^2$ carbon bonds, are predicted to have unique physical properties including isotropic planar elasticity with ultra-high in-plane stiffness and strength and Dirac fermions electronic structures (Novoselov et al., 2005; Zhang et al., 2005). Their thermal conductivity and mechanical stiffness/strength can reach the in-plane values of graphite and surpass most of the other engineering materials (Balandin et al., 2008; Gomez-Navarro et al., 2008; Lee et al., 2008; Stankovich et al., 2006). Due to its high stiffness and perfect lattice structure, graphene sheet has great potentials in making nano-electromechanical devices featuring resonant frequencies exceeding 100 MHz (Bunch et al., 2007; Chen et al., 2009; Robinson et al., 2008) and quality-factors up to $10^4$ (Chen et al., 2009). Recently the rapid development of graphene synthesis and transfer techniques (Lee et al., 2010; Li et al., 2009; Wei et al., 2008) makes mass production of graphene-based nano-devices more feasible.

Resonant frequency is extremely important for a resonant device, as many sensors based on mechanical resonators just utilize the change in their resonant frequencies to detect exotic signals such as adsorption, contamination and electromagnetic field (Ekinci and Roukes, 2005; Ekinci et al., 2004; Huang et al., 2003; Poncharal et al., 1999). For optimized design of graphene nano-strip-based electromechanical devices, it is necessary to understand the transverse vibration dynamics and corresponding resonant frequency of the beam structure, while taking into account that the out-of-plane deformation and localized ripple of the graphene sheet also modify the charge distribution and its electron transport properties (Bolotin et al., 2008; Gibertini et al., 2010).

Graphene is only one atom thick, the thinnest elastic material in the nature. This lets graphene sheets have some exceptional mechanical properties (Lee et al., 2008; Meyer et al., 2007). The in-plane stiffness and strength of graphene are exceptionally high (Gomez-Navarro et al., 2008; Lee et al., 2008). But on the other hand the graphene can be very soft for out-of-plane deformation. Ripples are intrinsic feature of suspended graphene sheets (Meyer et al., 2007). A regular one- or two-dimensional ripple texture can be easily generated in a suspended graphene film by different thermal strains in the graphene and in the substrate (Bao et al., 2009). Moreover, a supported graphene can follow the substrate to generate complicated morphology (Li and Zhang, 2010). The mechanical properties of a single layer graphene can be described by continuum shell models (Arroyo and Belytschko, 2002; Huang et al., 2006; Wang et al., 2005; Yakobson et al., 1996). In order to describe the mechanical properties of multilayer graphene based materials, such as multi-walled carbon nanotube and multilayer graphene sheets, the interlayer van der Waals interaction has been incorporated into the continuum shell models of the single layer graphene (He et al., 2005; Liew et al., 2006; Ru, 2001). However the existing continuum models for multilayer graphene fail to consider the shear energy between the graphene layers. As the very small thickness of the graphene leads to a very small bending rigidity, the interlayer shear energy may play a very important role in bending and vibration of multilayer graphene sheets and must be considered in the related continuum models.

In multilayer graphene sheets, unlike in conventional bulk materials often used in microelectromechanical devices, the interlayer van der Waals binding is very weak. Self-retracting behavior was discovered in slided graphite microflakes, indicating an extremely low resistance against shear and representing a new sliding-retracting motion mode (Zheng et al., 2008). Similar self-retracting behavior was also found in multi-walled carbon nanotubes with the interlayer shear strength estimated as 0.3 - 0.48 MPa (Cumings and Zettl, 2000; Yu et al., 2000a; Yu et al., 2000b). This unique feature not only provides a highly mobile interface between neighboring graphene sheets, which can be purposefully designed for use in mechanical devices, but also has dramatic effects on the physical and chemical properties of multilayer graphene sheets. Earlier works studying the dynamical behaviors of graphene resonators also did not pay serious attention to this issue (Bunch et al., 2007; Chen et al., 2009; Robinson et al., 2008). In general, effects of the interlayer shear on the vibrational behavior of multilayer graphene beam are still not to be clarified. In order to investigate the effects of interlayer shear on resonant frequency without axial in-plane tension involved, we will study cantilever beams made from various numbers of layers, through molecular dynamics (MD) simulations and theoretical analysis.

## 2. Molecular Dynamics Simulations

In our MD simulations Dreiding force-field is taken into account, which describes both the intra-layer and inter-layer interactions between carbon atoms (Mayo et al., 1990) and has been validated for characterizing the structural and mechanical properties of carbon nanostructures (Guo et al., 1991). The MD code GROMACS (Lindahl et al., 2001) is used. Graphene nanostrips investigated here have rectangular cross-sections with a fixed width of 2 nm but different heights measured by their layer number $n$, and have strip lengths ranging from 8 to 16 nm. We apply a displacement of 1 nm to the free end of each cantilevered nano-strip at the beginning of the simulation to initialize a transverse vibration. The atomic structures of the multilayer graphene nano-strips at initiation are illustrated in Fig. 1 (a), showing a remarkable shear at the free end of each studied multilayer graphene cantilever. The free end transversal motion of the strips is tracked, as plotted in Fig. 1 (b). We see that the vibration sustains without significant dissipation in the period of one nanosecond. We also see that the displacement-time curve deviates from ideal sinusoidal function, featuring numerous corrugations that result from local ripples of the graphene sheets. This phenomenon is unique for fewer-layer graphene sheets in which higher-order beam-like modes with much shorter wavelength are immediately excited in addition to the first-order mode as initialized. The Fast Fourier Transformation (FFT) results are shown in Fig. 1 (c), clearly indicating domination of the fundamental resonance.

From Fig. 1 (c) we can see that the resonant frequencies shift not much when the graphene layer number exceeds two. To see more clearly the dependence upon layer number $n$, we plot these frequencies as black solid triangles in Fig. 2 (a). As shown by the black dashed line, the simulated frequencies can be least-square fitted into a relationship $f = f_1 + \Delta f\,(n-1)/n$, in which $f_1 = 3.7$ GHz is the simulated frequency of the monolayer cantilever and $\Delta f$ = 7.7 GHz the fitted constant, with an excellent agreement. This result reveals that increasing layer number has no significant influence to the resonant frequency, which with the increased layer-number tends to a saturation frequency $f_{sa} = f_1 + \Delta f$ = 11.4 GHz.

We have also investigated the frequency-length relationship of multilayer graphene nano-strips by using MD simulations. Resonant frequencies are calculated for nano-strips with lengths, $L$, ranging from 8 to 16 nm. The black solid triangles plotted in Fig. 2 (b) represent the simulated results of three-layer graphene nano-strips, and the black dashed curve is the least-square fitting of the results - a power law $f = a(10h/L)^b$ with two fitted parameters $a$ = 40.1 GHz and $b$ = 1.36, which shows an excellent agreement, where $h$ = 0.335 nm is the interlayer spacing. Similar scaling laws are found valid for graphene nano-strips with 4 through 6 layers, in which $b$ = 1.34, 1.33, 1.33, respectively.

**3. Analytical Models**

*3.1 Euler-Bernoulli-Model*

Due to their large length-to-thickness ratios, typical thin beams are very well described by classical Euler-Bernoulli beam model (EBM), which ignores the influence of shear. For a cantilever beam with a uniform rectangular cross-section made of a homogeneous linear elastic material, its resonant frequencies are known to be

$$f = \frac{\beta^2}{2\pi L^2}\sqrt{\frac{E}{12\rho}}H, \qquad (1)$$

where $E$ and $\rho$ are Young's modulus and mass density of the material, $H$ the thickness, and $\beta$ the resonant mode parameter that is one of the solutions of equation

$$1 + \cosh\beta \cos\beta = 0, \qquad (2)$$

giving the values of $\beta$ = 1.875, 4.694, 7.855,…, for the leading resonant modes.

If EBM is applied to our studied $n$-layer graphene beams, we have $H = nh$, where

$h = 0.335$nm is the interlayer spacing. (1) can be reformulated into the following form:

$$f = \frac{\beta^2}{2\pi L^2}\sqrt{\frac{D_{bend}}{\rho h}}n = f_{mono}n, \tag{3}$$

with

$$f_{mono} = \frac{\beta^2}{2\pi L^2}\sqrt{\frac{D_{bend}}{\rho h}}, \tag{4}$$

where $D_{bend}$ is the bending rigidity of a monolayer graphene of unit width, and $f_{mono}$ the resonant frequency of a monolayer graphene cantilever beam. Thus, EBM gives a scaling law $f \propto L^{-2}$ and linear dependence upon the layer number $f \propto n$, which are both severely departed from respective MD simulations results $f \propto L^{-1.36}$ and $f - f_{mono} \propto (n-1)/n$. To view the departures directly, we also plot the EBM predictions of $f$ versus $n$ and $L$ in Fig. 2. Here, bending rigidity $D_{bend}$ of the monolayer graphene sheet with unit width equals $2.72 \times 10^{-19}$ kgm$^2$s$^{-2}$, which is calculated through rolling monolayer graphene sheets into cylindrical tubes with various radius $R$, and equating the exceeding energy to the elastic energy by shell model with the same deformation, while using the same interaction potential of Dreiding force-field model (Mayo et al. 1990).

*3.2 Multi-beam Shear Model*

In order to obtain some insights into this remarkable difference, we make such an analysis as introduced below. For the single-walled carbon nanotube, Yakobson et al. (Yakobson et al., 1996; see also Wang et al., 2005; Chang and Gao, 2003) suggested that Young's modulus and thickness should be defined consistently as 5.5 TPa and 0.066 nm to enable the continuum shell model to accurately characterize the elastic response. It is notable that thickness 0.066 is one-fifth of the intuitive definition of graphene sheet thickness (interlayer distance in graphite or multi-walled carbon nanotubes). Thus, a more reasonable model for a multilayer graphene nano-strips should be a structure of multi-beams. Indeed, by carefully looking at the deformation and motion of the multilayer graphene cantilever nano-strips during vibration (see, for example, Fig. 1 (a)) we discover that the weak inter-layer van der Waals interaction cannot maintain the registry of carbon atoms in adjacent layers, and consequently the cantilevers cannot be considered as an integrated solid. Instead, we must incorporate the inter-layer shear into our analysis for a more accurate estimation of the resonant frequencies of the multilayer graphene cantilever nano-strips. Here we propose a new model, which can be named as multi-beam shear model (MBSM). We firstly assume that every graphene layer behaves as a cantilever beam and then introduce a potential energy term accounting for the shear between neighboring graphene layers. From the MD simulation results we find that vibrations of all graphene layers are coherent, i.e. their displacements are almost the same. Thus we further make a simplification that every beam has the same transverse displacement $w$, and hence the same interlayer shear $w'$, where superscript ' denotes spatial derivative along $x$ direction in the beam contour. Fig. 3 gives an schematical illustration of this model, in which the vertical massless rigid bars keep all the beams with the same $w$ and the oblique massless springs establish proper interlayer shear elasticity, $G$. The component beams are assumed to be inextensible because in-plane Young's modulus is too large compared to interlayer shear modulus (Kelly, 1981). The total potential of this multi-beams system in equilibrium under a uniformly distributed transverse load $q$ is thus equal to:

$$\Pi(w) = n \frac{D_{\text{bend}}}{2} \int_0^L (w'')^2 dx + (n-1) \frac{D_{\text{shear}}}{2} \int_0^L (w')^2 dx - \int_0^L qw\,dx, \qquad (5)$$

where $D_{\text{shear}}$ is the graphene interlayer shear rigidity of unit width. $D_{\text{shear}} = Gh$, where $G$ is interlayer shear modulus of graphene, and $h$ interlayer space of multi-layer graphene. The variation of $\Pi$ with respect to $w$ is derived as:

$$\begin{aligned}\delta \Pi(w) &= nD_{\text{bend}} \int_0^L w'' \delta w'' dx + (n-1) D_{\text{shear}} \int_0^L w' \delta w' dx - \int_0^L q \delta w\,dx \\ &= \int_0^L [nD_{\text{bend}} w'''' - (n-1) D_{\text{shear}} w'' - q] \delta w\,dx \\ &\quad + nD_{\text{bend}} w'' \delta w' \big|_0^L - [nD_{\text{bend}} w''' - (n-1) D_{\text{shear}} w'] \delta w \big|_0^L\end{aligned} \qquad (6)$$

Requiring $\delta \Pi = 0$ yields the basic differential equation:

$$nD_{bend}w'''' - (n-1)D_{shear}w'' - q = 0, \qquad (7)$$

and substituting boundary conditions at the clamped end $\delta w(0) = 0$ and $\delta w'(0) = 0$ yields the boundary conditions at the free end:

$$\begin{aligned} & w''(L) = 0, \\ & D_{bend}w'''(L) - \frac{n-1}{n}D_{shear}w'(L) = 0. \end{aligned} \qquad (8)$$

Because of the interlayer shear, transverse force boundary condition (8)$_2$ at the free end is different from that of a classical Euler-Bernoulli cantilever.

To investigate free vibration of the system, the distributed load $q$ is withdrawn and the inertial force $-n\rho h \ddot{w}$ comes into effect, and the control equation becomes:

$$D_{bend}w'''' - \frac{n-1}{n}D_{shear}w'' + \rho h \ddot{w} = 0. \qquad (9)$$

For a harmonic vibration $w(x,t) = W(x) \sin\omega t$, it yields

$$\frac{d^4W}{d\xi^4} - \frac{n-1}{n}\frac{D_{shear}L^2}{D_{bend}}\frac{d^2W}{d\xi^2} - \frac{\rho h L^4}{D_{bend}}W = 0, \qquad (10)$$

where $\xi = x/L$ is the dimensionless coordinate. The general solution of Eq. (10) can be expressed as:

$$W = A_1 \cosh \beta_+\xi + A_2 \cos \beta_-\xi + A_3 \beta_- \sinh \beta_+\xi + A_4 \beta_+ \sin \beta_-\xi, \qquad (11)$$

where

$$\begin{aligned} \beta &= \sqrt{\sqrt{\frac{\rho h}{D_{bend}}}\omega L}, \\ \eta &= \frac{n-1}{2n}\frac{D_{shear}h^2}{D_{bend}}(\frac{L}{h})^2, \\ \beta_\pm &= \sqrt{\sqrt{\beta^4 + \eta^2} \pm \eta}. \end{aligned} \qquad (12)$$

By substituting the boundary conditions at the clamped end $W(0) = 0$ and $W'(0) = 0$ into general solution (11), we get relations $A_2 = -A_1$ and $A_4 = -A_3$. These relations, together with general solutions (11), are substituted into boundary Eqs. (8), leading to the following boundary equations:

$$(\beta_+^2 \cosh\beta_+ + \beta_-^2 \cos\beta_-)A_1 + \beta_-\beta_+(\beta_+ \sinh\beta_+ + \beta_- \sin\beta_-)A_3 = 0,$$
$$[(\beta_+^3 \sinh\beta_+ - \beta_-^3 \sin\beta_-) - 2\eta(\beta_+ \sinh\beta_+ + \beta_- \sin\beta_-)]A_1$$
$$+ [\beta_-\beta_+(\beta_+^2 \cosh\beta_+ + \beta_-^2 \cos\beta_-) - 2\eta\beta_+\beta_-(\cosh\beta_+ - \cos\beta_-)]A_3 = 0$$
(13)

To obtain nonzero solutions of $A_1$ and $A_3$, the determinant of the coefficient eigen-matrix of (13) must be zero, viz.

$$\beta_+^4 + \beta_-^4 - 2\eta(\beta_+^2 - \beta_-^2) - \beta_+\beta_-(\beta_+^2 - \beta_-^2 - 4\eta)\sin\beta_- \sinh\beta_+$$
$$+ 2\eta(\beta_+^2 - \beta_-^2)\cos\beta_- \cosh\beta_+ + 2\beta_+^2\beta_-^2 \cos\beta_- \cosh\beta_+ = 0$$
(14)

Using the following identities:

$$\beta_+^2 - \beta_-^2 = 2\eta,$$
$$\beta_+^2 \beta_-^2 = \beta^4,$$
$$\beta_-^4 + \beta_+^4 = 4\eta^2 + 2\beta^4.$$
(15)

we can further simplify Eq. (14) into the following eigen-equation:

$$1 + (1 + 2\frac{\eta^2}{\beta^4})\cosh\beta_+ \cos\beta_- + \frac{\eta}{\beta^2}\sinh\beta_+ \sin\beta_- = 0 \quad (16)$$

Whenever $\eta = 0$, Eq. (16) degenerates into Eq. (2) - the eigen-equation for EBM.

*3.3 Comparsion of MBSM and MD simulations*

Any solution $\beta$ of Eq. (16) corresponds to a resonant frequency $f = \omega/2\pi$ as seen from Eq. (12)$_1$. However, unlike solutions of Eq. (2) that are independent of $L$, a solution $\beta$ of Eq. (16) will depend upon $L$ through $\eta$, $\beta_+$, and $\beta_-$, which are related to $L$ by Eqs. (12)$_{2,3}$. Thus a scaling law of $f$ will be resulted, which is different from $f \propto L^{-2}$. There is still an undetermined parameter involved in MBSM, namely $D_{shear}/D_{bend}$ - the ratio of interlayer shear rigidity $D_{shear}$ to bending rigidity $D_{bend}$ of monolayer graphene. We evaluate bending rigidity $D_{bend}$ through substituting the foundamental resonant frequency of monolayer graphene cantilever obtained in our MD simulation into the form of EBM frequency $f_{mono} = \frac{1.875^2}{2^1 l^2}\sqrt{\frac{D_{bend}}{\rho h}}$. This gives $D_{bend}$ = 3.42×10$^{-19}$ kgm$^2$s$^{-2}$.

To determine interlayer shear rigidity $D_{shear}$, which is defined as the product of interlayer shear modulus $G$ and interlayer space $h$, we do a MD numerical experiment, in which a small flake of single-layer graphene is sliding on a large single-layer graphene substrate and van der Waals forces between all atoms in the flake and the graphene substrate are calculated. The sliding direction is the same as what we have observed in the vibration of multilayer graphene nano-strips, i.e. armchair direction. We use Lennard-Jones formula to represent the interlayer van der Waals interaction, with the formula and parameters all the same as used before for the inter-layer interaction of Dreiding force-field (Mayo et al., 1990). In our calculation, shear strain is defined as $\gamma = s/h$, where $s$ is the displacement of the sliding graphene flake and $h$ the graphene interlayer space. The calculated relationship between shear stress and strain is plotted in Fig. 4, from which we estimate shear modulus $G$ = 0.25 GPa by linear fitting the curve in the range of small strains. With interlayer space $h$ = 0.34 nm, we get the shear rigidity of unit width $D_{shear}$ = 0.085 kgs$^{-2}$.

Consequently we can calculate the values of $\eta$ for different layer numbers according to Eq. (12)$_2$, among which $\eta$ =6.2 for two-layers graphene nano-strip and the upper bound $\eta$ = 12.4 for infinite graphene layers. According to Eqs. (12) and Eq. (16), the large value of $\eta$ will largely influence the resonant mode parameter $\beta$, meaning that the interlayer shear in multi-layer graphene cantilever nano-strips has very large influence on their resonant frequencies.

Subsequently, we calculate the resonant frequencies of multi-layer graphene cantilever nano-strips by numerically solving Eq. (16) while using the parameters determined from the MD simulation. Thus calculated dependence of the resonant frequencies on the number of layers is shown in Fig. 2 (a) with red circles, showing a very good agreement between our MBSM predictions and the direct MD simulations.

The dependence of the resonant frequencies on the length of the nano-strips is also shown in Fig. 2 (b) with red circles, also agreeing very well with the MD simulations. Because MBSM has properly considered the role of interlayer shear energy during vibration of multi-layer grapheme nano-strips, it makes very good predictions consistent with MD simulations.

**4. Closing Remarks**

In this work we find through molecular dynamics simulations that the resonant frequencies $f$ of multilayer graphene cantilevers depend upon beam length $l$ and layer number $n$ in a manner very different from classical *Euler-Bernoulli-Model* prediction $f \propto nl^{-2}$, which is generally valid for thin beams. Our MD simulations can be perfectly described in terms of a scaling law $f \propto l^{-b}$ with $b = 1.36, 1.34, 1.33, 1.33$ as $n = 3,\ldots,6$, respectively and a layer number dependence $f - f_{mono} \propto (n-1)/n$, where $f_{mono}$ is the resonant frequency of a monolayer graphene cantilever.

It is also observed from MD simulated instant configurations of the multilayer graphene cantilevers during vibrations that remarkable interlayer shear happens and all the layers have almost the same deflections. This is consistent with the known great contrast between intralayer Young's modulus and interlayer shear modulus. Based on this observation, we propose a multi-beam shear model (MBSM) which gives excellent predictions agreeing well with MD simulations. The proposed MBSM contains only two material parameters, bending rigidity $D_{bend}$ of monolayer graphene sheet and interlayer shear rigidity $D_{shear}$. In our practice, $D_{bend}$ is obtained by combing MD simulated resonant frequency of monolayer graphene nano-strip with EBM frequency relation $f_{mono} = \dfrac{1.875^2}{2\pi l^2}\sqrt{\dfrac{D_{bend}}{\rho h}}$; $D_{shear}$ is obtained by a MD simulayion, in which a graphene flake slides on a graphene substrate with the same interaction potential. MBSM does not need any fitting parameter but makes predictions agreeing excellently with direct MD simulations, implying that this model has captured the intrinsic properties dominanting vibration of multilayer graphene nano-strips.

In addition, MBSM has also been tested for predicting resonating behaviors of two-end clamped multilayer graphene beams with similarly good effects.

An open problem should be addressed here. In this work we find static bending rigidity of a single layer graphene sheet is different from its dynamic bending rigidity. The static bending rigidity is $2.72 \times 10^{-19}$ $kgm^2s^{-2}$, which is calculated through rolling monolayer graphene sheets into cylindrical tubes with various radius $R$, and equating the exceeding energy to the elastic energy by shell model with the same deformation. While the dynamic bending rigidity is $3.42 \times 10^{-19}$ $kgm^2s^{-2}$, 1.26 times of the static rigidity, which is calculated by substituting the MD predicted first resonant frequency of monolayer graphene nano-strip into EBM frequency formula. Similar result has not been seen from the literature yet. One possible explanation is that during vibration a monolayer grapahene may be locally curved due to temperature and thus generates higher bending rigidity. Further work is under way to systematically study this dynamic effect.

**Acknowledgements**

Q.-S. Zheng acknowledges the financial supports from NSFC (Grant No. 10832005), the National Basic Research Program of China (Grant No. 2007CB936803), and the National 863 Project (Grant No. 2008AA03Z302).

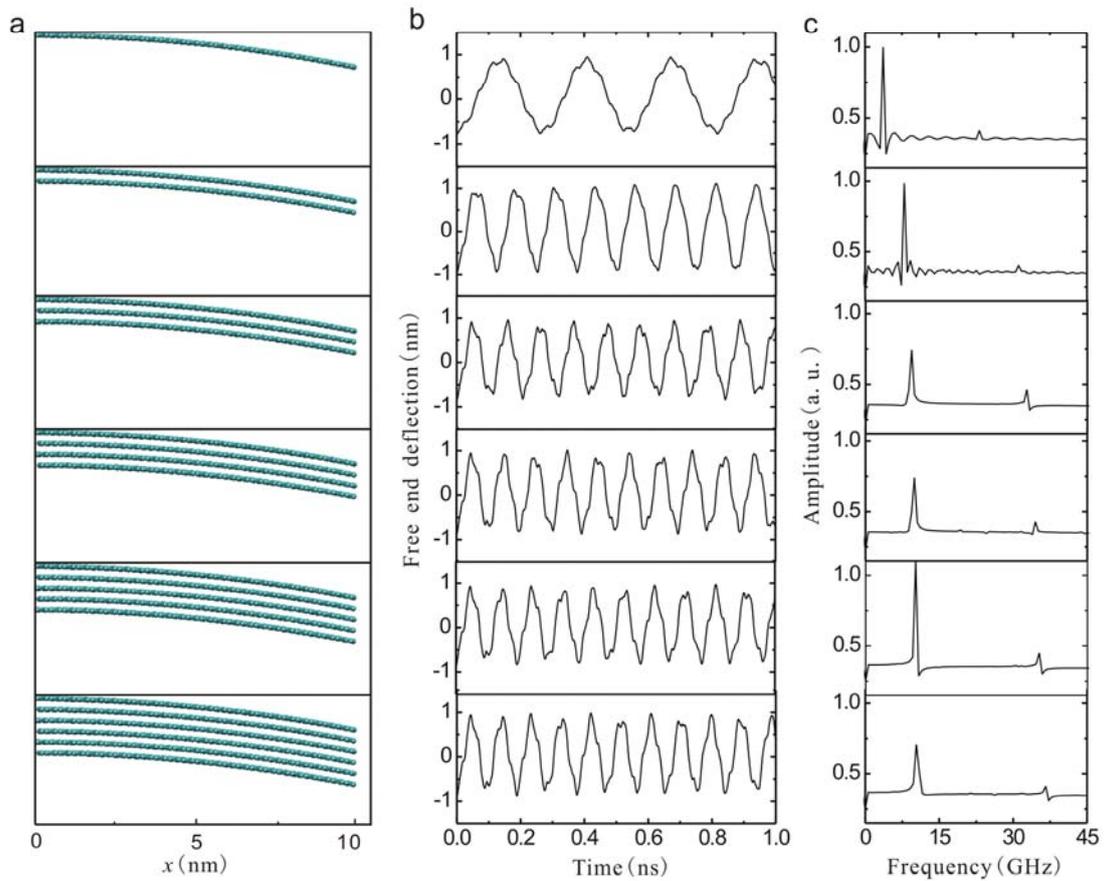

*Figure 1* The transversal vibration motion of the graphene cantilever nano-strips. (a) Atomic structures of the cantilevers consisting of one through six layers, with a width of 2 nm and length of 10 nm. Initial amplitude of the first-order vibration is 1 nm. (b) Free-end displacements of the corresponding graphene nano-strips. (c) Resonant frequencies of the corresponding graphene cantilever nano-strips resulted from Fast Fourier Transformation of (b).

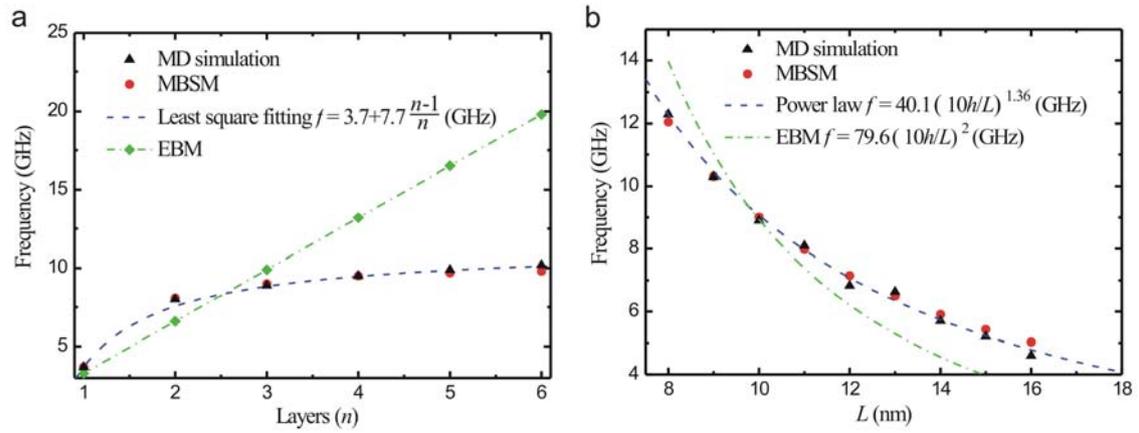

*Figure 2* The resonant frequencies of graphene cantilever nano-strips with different layers and different length. (a) Frequencies of graphene cantilever nano-strips consisting of one through six layers. Different results are from empirical molecular dynamics (MD) simulation (black triangles), its least square fitting (black dashed line), Euler-Bernoulli beam model (EBM) prediction (green squares), and multi-beam shear model (MBSM) prediction (red circles). (b) Scaling laws of the resonant frequencies of three-layer graphene cantilever nano-strips. Different results are from MD simulation (triangles), its power-law fitting (black dashed line), MBSM prediction (circles) and EBM prediction (green dash-dotted line).

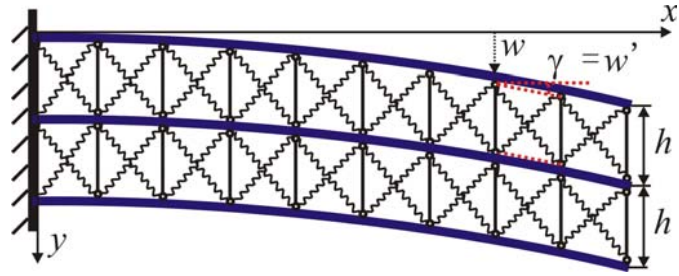

*Figure 3* Schematic illustration of MBSM model. The vertical massless rigid bars keep every beam having the same deflection and the oblique massless springs establish interlayer shear, where h is interlayer space, w is deflection of every beam, and w' interlayer shear strain as indicated with dashed red line.

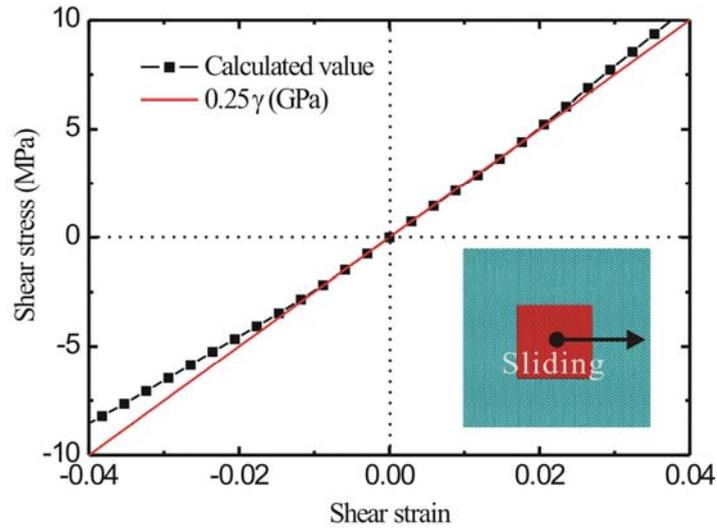

*Figure 4 Relationship between interlayer shear stress and strain of graphene layers, calculated by summing the van der Waals forces between a graphene flake and a substrate while the flake is sliding with respect to the substrate, with Lennard-Jones formula used. Linear fitting to the numerical results gives a shear modulus G = 0.25 GPa.*